\documentclass[%
reprint,
amsmath,amssymb,
aip,
jcp
]{revtex4-1}

\usepackage{graphicx}
\usepackage{dcolumn}
\usepackage{bm}
\usepackage{hyperref}
\usepackage[english]{babel}
\usepackage{color}
\usepackage{epstopdf}

\begin{document}

\title{The influence of random pinning on the melting scenario of two-dimensional soft-disk systems}

\author{E. A. Gaiduk}
\affiliation{Institute for High Pressure Physics, Russian Academy
of Sciences, 108840, Troitsk, Moscow, Russia
}
\author{Yu. D. Fomin}
\affiliation{Institute for High Pressure Physics, Russian Academy
of Sciences, 108840, Troitsk, Moscow, Russia
}
\author{E. N. Tsiok}
\affiliation{Institute for High Pressure Physics, Russian Academy
of Sciences, 108840, Troitsk, Moscow, Russia
}
\author{V. N. Ryzhov}
\affiliation{Institute for High Pressure Physics, Russian Academy
of Sciences, 108840, Troitsk, Moscow, Russia
}

\begin{abstract}
We present the results of a computer simulation study of the
melting scenario of two-dimensional soft-disk systems with
potential $U(r)=\varepsilon(\sigma/r)^n$, $n=12$ and $n=1024$,
both in the presence of random pinning and without it. The melting
parameters have been determined from analysis of the behavior of
equations of state, correlation functions of the orientational and
translational order parameters, Young modulus and renormalization
group equations. The transition points obtained from these
criteria are in good agreement. We have shown that the systems
melted in two stages - first-order hexatic phase to liquid
transition and the continuous Berezinskii-Kosterlitz-Thouless type
crystal-hexatic phase transition. Random pinning widened the
hexatic phase, but left the melting scenario unchanged.
\end{abstract}

\pacs{61.20.Gy, 61.20.Ne, 64.60.Kw}
\date{\today}

\maketitle

\section{Introduction}

The problem of two-dimensional (2D) melting has been of great
interest for more than 40 years already. However, even now there
is no microscopic theory to satisfactorily describe the nature of
two-dimensional melting, which remains among the oldest unsolved
problems in condensed matter physics. In spite of a large number
of publications there are still many controversies. While in three
dimensional (3D) space melting is always first-order phase
transition, there are several different scenarios of melting in 2D
\cite{ufn1,ufn2}. The main reason of this difference is related to
strong fluctuations in 2D compared to the 3D case. Peierls
\cite{p1,p2} and Landau \cite{lan} and later Mermin \cite{mermin}
demonstrated that in 2D crystals there was no long-range
translational order. The translational order of 2D crystals is in
fact quasi long-range, i.e. the correlation functions decay
algebraically. However, there is another type of long-range order
in 2D crystals: orientational order which is the order in bonds
between a particle and its nearest neighbors \cite{mermin}. At
high temperatures a 2D system is in a usual state of isotropic
liquid.

The Berezinskii-Kosterlits-Thouless-Halperin-Nelson-Young
(BKTHNY)\cite{b1,kosthoul73,halpnel1,halpnel2,halpnel3} theory of
2D melting is the most common. Within the framework of this theory
2D melting takes place in two stages. At the first stage a crystal
transforms into a so called hexatic phase via continuous phase
transition. This transformation is caused by dissociation of
coupled dislocation pairs. Below we will refer to the temperature
of dislocation pairs dissociation as $T_m$. This dissociation
leads to a complete break of translational order: while in 2D
crystals the translational order is quasi long-range, in the
hexatic phase it becomes short-range. At the same time the
orientational order changes from long-range in crystals to quasi
long-range in the hexatic phase.

Dislocations themselves can be considered as bounded pairs of
disclinations. These pairs also dissociate at higher temperature
$T_i$ where the system transforms into isotropic liquid. Within
the framework of the BKTHNY theory this transition is also a
continuous one. The BKTHNY theory was confirmed by some
experimental studies (see, for instance,
\cite{keim1,zanh,keim2,keim3,keim4} where systems of colloids in
magnetic fields are studied).

However, 2D melting can also occur through first-order phase
transition alone. In Ref. \cite{chui83} it was shown that at low
energy of the dislocation core the dissociation of bound
dislocation pairs was preempted by the proliferation of grain
boundaries leading to the first-order melting transition.
First-order transition takes place at dislocation core energy
$E_c$ less than $2.84 T_m$, where $T_m$ is obtained from equation
(\ref{tm}) (see below). A similar mechanism leading to first-order
transition was discussed in Refs.\cite{ryzhovTMP,ryzhovJETP}.
Since dislocation is a disclination dipole a bound dislocation
pair can be considered as a disclination quadrupole. As shown in
papers \cite{ryzhovTMP,ryzhovJETP} there is a critical value of
disclination core energy below which the dissociation of
disclination quadrupoles leads to the appearance of free
disclinations and the transition of first-order to isotropic
liquid. The dependence of the melting scenarios on the form of the
potential was studied in Refs. \cite{rto1,rto2,RT3,RT4} within the
framework of the density functional theory of crystallization. A
unified approach to the description of first-order and BKT type
melting of the solid phase within the framework of Landau's theory
of phase transitions was proposed in Refs. \cite{RT1,RT2}.

Although the BKTHNY theory does not depend on the interaction
potential, many experiments and computer simulations show that
such dependence exists. It was widely believed that systems with
short-range interaction potentials melted via first-order phase
transition, while long-range interactions, for example, the
Coulomb one, led to the BKTHNY melting scenario \cite{ufn1}.

However, even for the simplest system of hard spheres the results
are contradictory (see, for example, \cite{binderPRB,mak,binder,nez}).
This contradiction was partly resolved in the series of papers
\cite{foh1,foh2,foh3}, where based on the results of computer
simulations it was proposed that melting took place as a two stage
process, but while transition from crystal to hexatic was a
continuous one, the second transition from hexatic to isotropic
liquid was a first-order transformation. This new melting scenario
was experimentally confirmed in Ref. \cite{hsn}.

In particular, Ref. \cite{foh4} presents a study of melting of
soft disks (a system of particles interacting via inverse power
potential $U \sim 1/r^n$) with a different n. It is shown that for
small $n<6$ the system melts in accordance with the BKTHNY theory,
while for $n>6$ the transition between the hexatic and liquid
phases becomes a first-order one, i.e. melting follows the third
scenario. Below we will call this scenario the BK one after
Bernard and Krauth who introduced it.

The melting of systems with core-softened potentials was studied
in Refs.  \cite{prest2,prest1,dfrt1,dfrt2,dfrt3,we2}. It was shown
that the system had a complex phase diagram with different crystal
structures, including a dodecagonal quasicrystal. Depending on the
position on the phase diagrams, the different melting scenarios
can take place, including first-order melting and the BK scenario.
In Refs. \cite{foh7,we3} computer simulations of a system of
purely repulsive soft colloidal particles interacting via the
Hertz potential and constrained to a two-dimensional plane were
presented. This potential can be a reliable model for a
qualitative description of behavior of soft macromolecules, such
as globular micelles and star polymers. A large number of ordered
phases was found, including the dodecagonal quasicrystal, and it
was shown that depending on the position on the phase diagram, the
system could melt through the first-order transition, in
accordance with the BKTHNY theory as well as according to the BK
scenario.

It should be noted that the region of stability of the hexatic
phase is usually rather narrow. As a result this region can be
almost invisible in experiments and computer simulations. In this
case transition from crystal to liquid may be mistaken for a
first-order one. In order to make the situation clearer it is
necessary to enlarge the region of stability of the hexatic phase.
One of the methods to do so is to introduce random pinning into
the system, i.e. to make some particles immobile and fix them in
some random positions, including interstitial lattice sites. As it
was shown theoretically \cite{nel_dis1,nel_dis2,dis3,dis4}, the
BKTHNY melting scenario persisted in this case. Random pinning
almost did not affect the transition between the hexatic phase and
isotropic liquid, while the solid phase could be completely
destroyed by high pinning fractions. In the latter case the
stability range of the hexatic phase increases due to moving the
border of the solid phase to higher densities on the equation of
state. This result was confirmed in experiments and simulations of
2D melting of super-paramagnetic colloidal particles with quenched
disorder \cite{keim3,keim4}. In Ref. \cite{we1} it was shown that
the BK scenario also persisted with random pinning. An even more
interesting result was obtained in Ref. \cite{dfrt4} where it was
demonstrated that random pinning could transform  first-order
melting into the BK scenario.

It should be specially mentioned that in the article by Qi and
Dijkstra \cite{foh6} the opposite case was considered where a
random fraction of particles was pinned on the regular lattice
sites of the crystal phase. In this case the pinned particles
stabilized the crystal lattice and consequently decreased the
range of the hexatic phase.

Below we will consider both the systems with and without random
pinning. We will also call systems without pinning pure systems.

In the present work we study the melting behavior of two soft disk
systems: with $n=12$ and with $n=1024$. In order to determine the
melting scenarios more accurately we introduce random pinning into
the system. It allows us to study the properties of the hexatic
phase itself, for example, the diffusion coefficient of the
hexatic phase. Moreover, we employ many different methods to
determine phase boundaries, including equations of state,
correlation functions, Young modulus and renormalization group
equations. This allows us to compare different approaches in order
to determine phase boundaries in 2D melting.

\section{System and Methods}

The soft disk system studied in this paper is defined by the interaction potential:

\begin{equation}
U(r)=\varepsilon \left( \frac{\sigma}{r} \right)^n.
\end{equation}
Parameters $\sigma$ and $\varepsilon$ can be used as the scales of
length and energy respectively. Based on these scales one can
construct the units of all other quantities of interest. For
instance, the density can be written as $\rho^*=\rho \cdot
\sigma^2$. Below we will use only these reduced units omitting
($^*$).

Two values of softness parameter $n$ are considered: $n=12$ and
$n=1024$. The latter $n$ is very close to the limiting case of
hard disks.

We simulate the system by means of molecular  dynamics method.
Mostly systems of 20000 particles are used. The system is
simulated in an NVT ensemble (constant number of particles N,
volume V and temperature T). Additional simulations in NVE
(constant number of particles N, volume V and internal energy E)
are performed to calculate the diffusion coefficient.

Firstly we perform simulations with a pure system. After that we
consider systems with $0.1 \%$ of particles pinned in random
positions. In this case we consider 10 different systems with
different initial positions of pinned particles. All results are
averaged over these 10 replicas.

Several methods are used to find phase boundaries. First of all we
analyze the equations of state. In the case of first-order phase
transition it should demonstrate the Mayer-Wood loop while in the
case of continuous transitions only a bend is observed. However,
from the equations of state one can only find whether the system
demonstrates a first-order phase transition, but it is impossible
to distinguish a simple first-order transition from the BK
scenario. In order to find the exact boundaries of the crystalline
phase we use the approach proposed in our recent publication
\cite{dfrt4}. As it was demonstrated in \cite{halpnel2}, the
orientational correlation function decays algebraically at the
boundary of the hexatic phase with isotropic liquid. Similarly,
the translational correlation function decays algebraically at the
boundary of the crystal and hexatic phases. In the presence of
random pinning the translational correlation function demonstrates
a bend. Algebraic decay is observed at large distances after the
bend \cite{dfrt4}.

The orientational order parameter is defined as
\begin{equation}
\Psi_6({\bf r_i})=\frac{1}{n(i)}\sum_{j=1}^{n(i)} e^{i
6\theta_{ij}}\label{psi6loc},
\end{equation}
where $\theta_{ij}$ is the angle between vector $\bf{ r}_{ij}$,
connecting the $i$-th and $j$-th particles and an arbitrary axis.
The sum includes all neighbors $n(i)$ of the $i$-th particle. The
nearest neighbors are determined by the Voronoi construction.

One can also define the global orientational order parameter
$\psi_6$, which is the average of $\Psi_6({\bf r_i})$  over the
whole system
\begin{equation}
\psi_6=\frac{1}{N}\left<\left<\left|\sum_i \Psi_6({\bf
r}_i)\right|\right>\right>_{rp}.\label{psi6}
\end{equation}

The second brackets $\left<({...})\right>_{rp}$ mean the averaging
over 10 replicas with random pinning.

The translational order parameter is defined as
\begin{equation}
\psi_T=\frac{1}{N}\left<\left<\left|\sum_i e^{i{\bf G
r}_i}\right|\right>\right>_{rp}, \label{psit}
\end{equation}
where ${\bf r}_i$ is the radius vector of the $i$-th particle and
{\bf G} is a reciprocal lattice vector.

The translational correlation function is defined as
\begin{equation}
G_T(r)=\left<\frac{<\exp(i{\bf G}({\bf r}_i-{\bf
r}_j))>}{g(r)}\right>_{rp}, \label{GT}
\end{equation}
where $r=|{\bf r}_i-{\bf r}_j|$, and $g(r)=<\delta({\bf
r}_i)\delta({\bf r}_j)>$  is a radial distribution function. In
the case of the crystalline phase without random pinning in the
limit $r \rightarrow \infty$ the translational correlation
function is $G_T(r)\propto r^{-\eta_T}$, where $\eta_T \leq
\frac{1}{3}$ \cite{halpnel1, halpnel2}.

The orientational correlation function $G_6(r)$ is defined in the same way:
\begin{equation}
G_6(r)=\left<\frac{\left<\Psi_6({\bf r})\Psi_6^*({\bf
0})\right>}{g(r)}\right>_{rp}, \label{g6}
\end{equation}
where $\Psi_6({\bf r})$ is the local orientational order parameter
(\ref{psi6loc}). The orientional correlation function decays
algebraically in the hexatic phase $G_6(r) \propto r^{-\eta_6}$
where $\eta_6 \leq \frac{1}{4}$ \cite{halpnel1,halpnel2,halpnel3}.
The limit of stability of the hexatic phase with respect to
isotropic liquid is determined from $\eta_6(T_i) = \frac{1}{4}$.

\section{Results and Discussion}

\begin{figure}
\includegraphics[width=8cm, height=8cm]{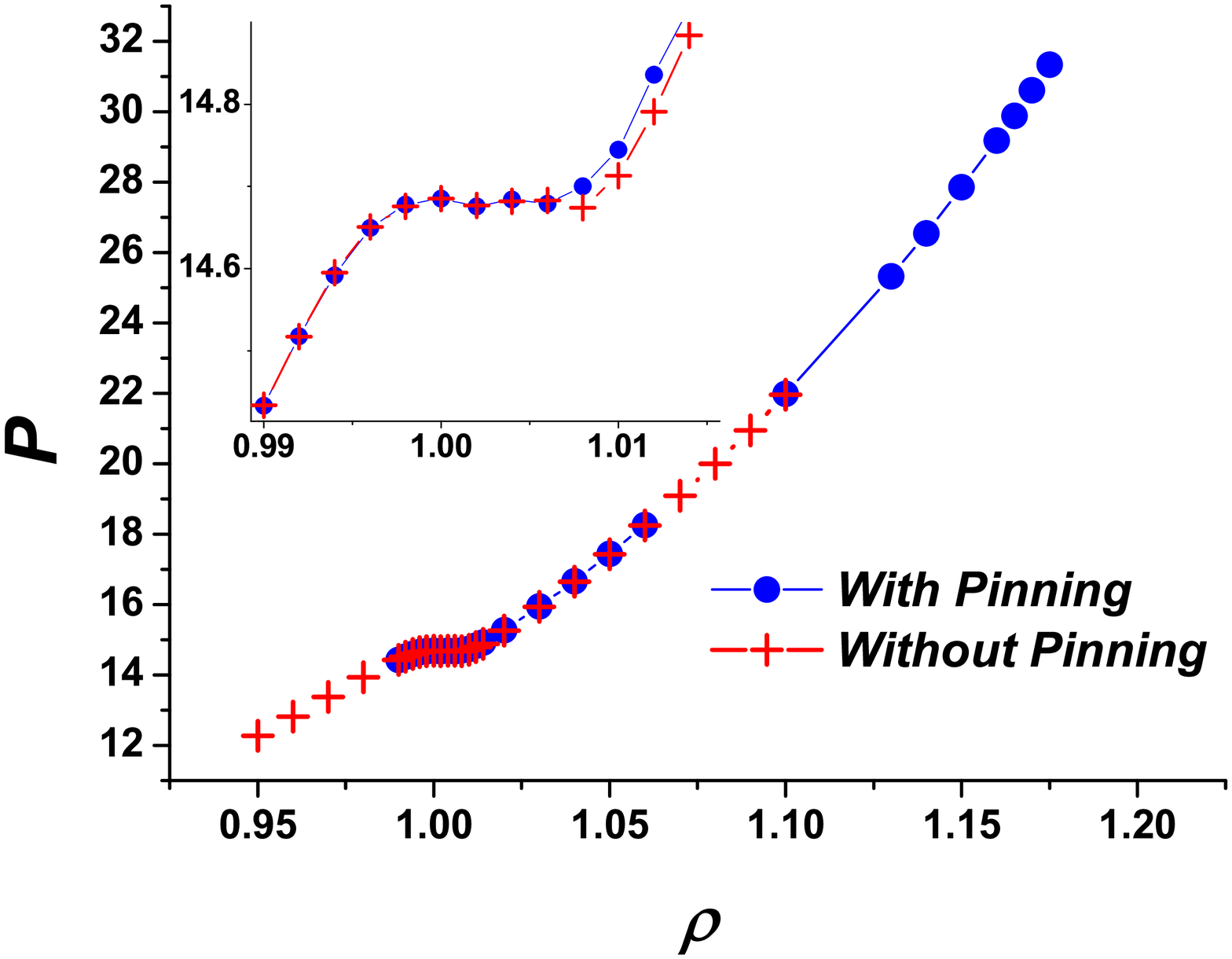}

\caption{\label{eos-n12} The equations of state of the system with
$n=12$ with and without random pinning at $T=1.0$. The inset
enlarges the region of the Mayer-Wood loop.}
\end{figure}

We start discussing the equation of state of the system with
$n=12$. Fig. \ref{eos-n12} shows the equations of state for both
the pure system and the system with random pinning.  At $T=1.0$
the Mayer-Wood loop is observed at densities around unity, i.e.
first-order phase transition takes place. However, it is necessary
to find  which phases are separated by this transition, whether it
is a simple first-order phase transition or the BK scenario. It
means that we need to determine which phase is located at the
right branch of the Mayer-Wood loop: crystal or hexatic. In order
to do this, we calculate the orientational order parameter, the
translational order parameter and their correlation functions.

\begin{figure}
\includegraphics[width=8cm, height=8cm]{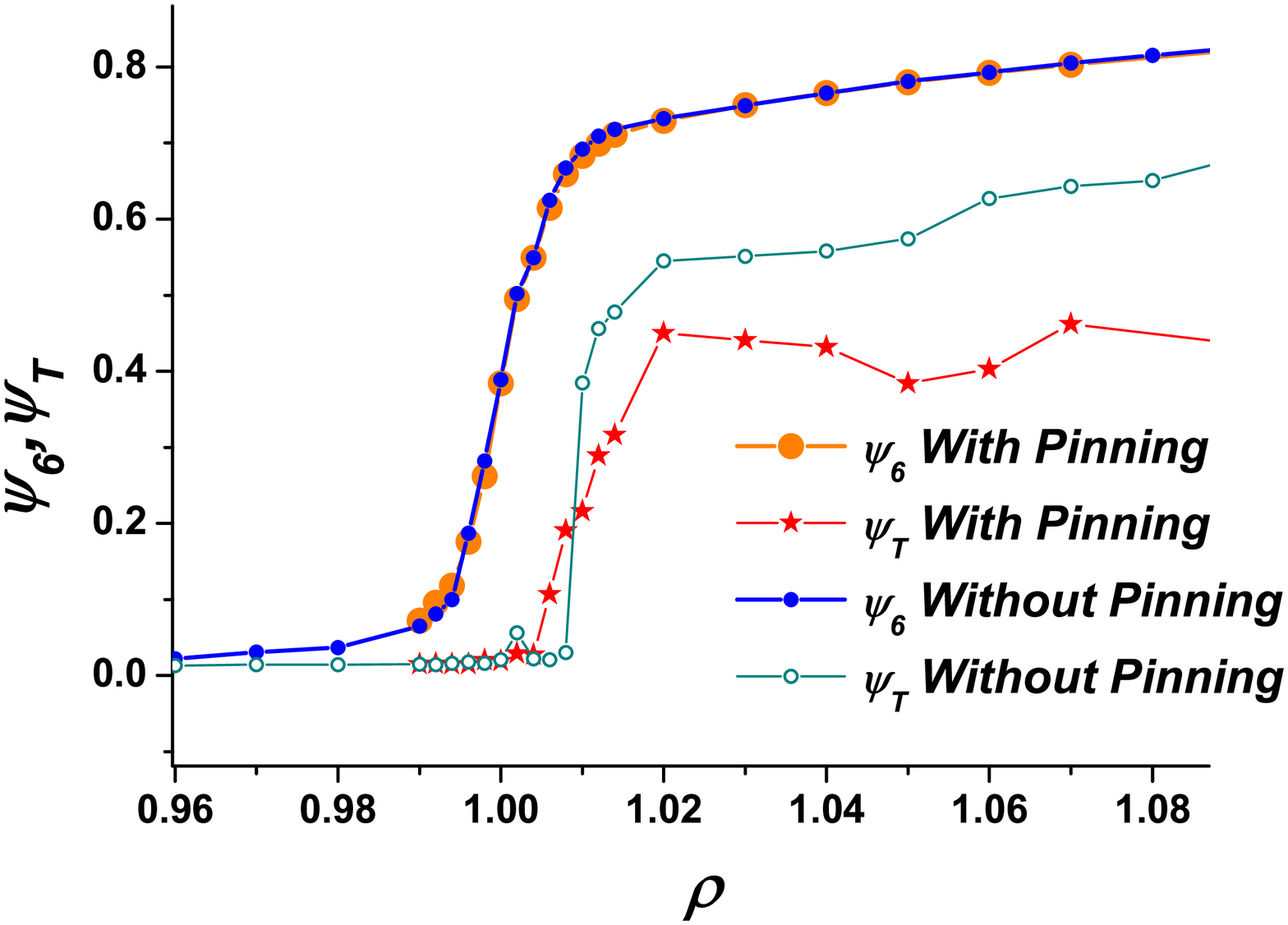}

\caption{\label{ops-12} The orientational and translational order
parameters of the $n=12$ system with random pinning and without
it.}
\end{figure}

Fig. \ref{ops-12} shows the orientational order parameter and the
translational order parameter both in the presence and absence of
random pinning along the $T=1.0$ isotherm. If there is no random
pinning the translational order parameter becomes zero at density
$\rho=1.008$, while the orientational order parameter becomes zero
at $\rho=0.99$. It means that there is a region where the system
is trans-lationally disordered while quasi long-range
orientational order is still present. This is consistent with the
definition of a hexatic phase.

Let us consider the order parameters of the system with random
pinning. The behavior of the orientational order parameter is the
same as in the pure system. However, the translational order
parameter behaves qualitatively different: starting from density
about $1.15$ there is a jump of the translational order parameter
at level about $0.45$. When the system enters the two-phase region
the translational order parameter vanishes (Fig. \ref{ops-12}). We
believe that the step corresponds to the region of the hexatic
phase. The finite value of the translational order parameter in
the hexatic phase should be related to some local ordering which
is not destroyed completely by random pinning.

\begin{figure}
\includegraphics[width=8cm, height=8cm]{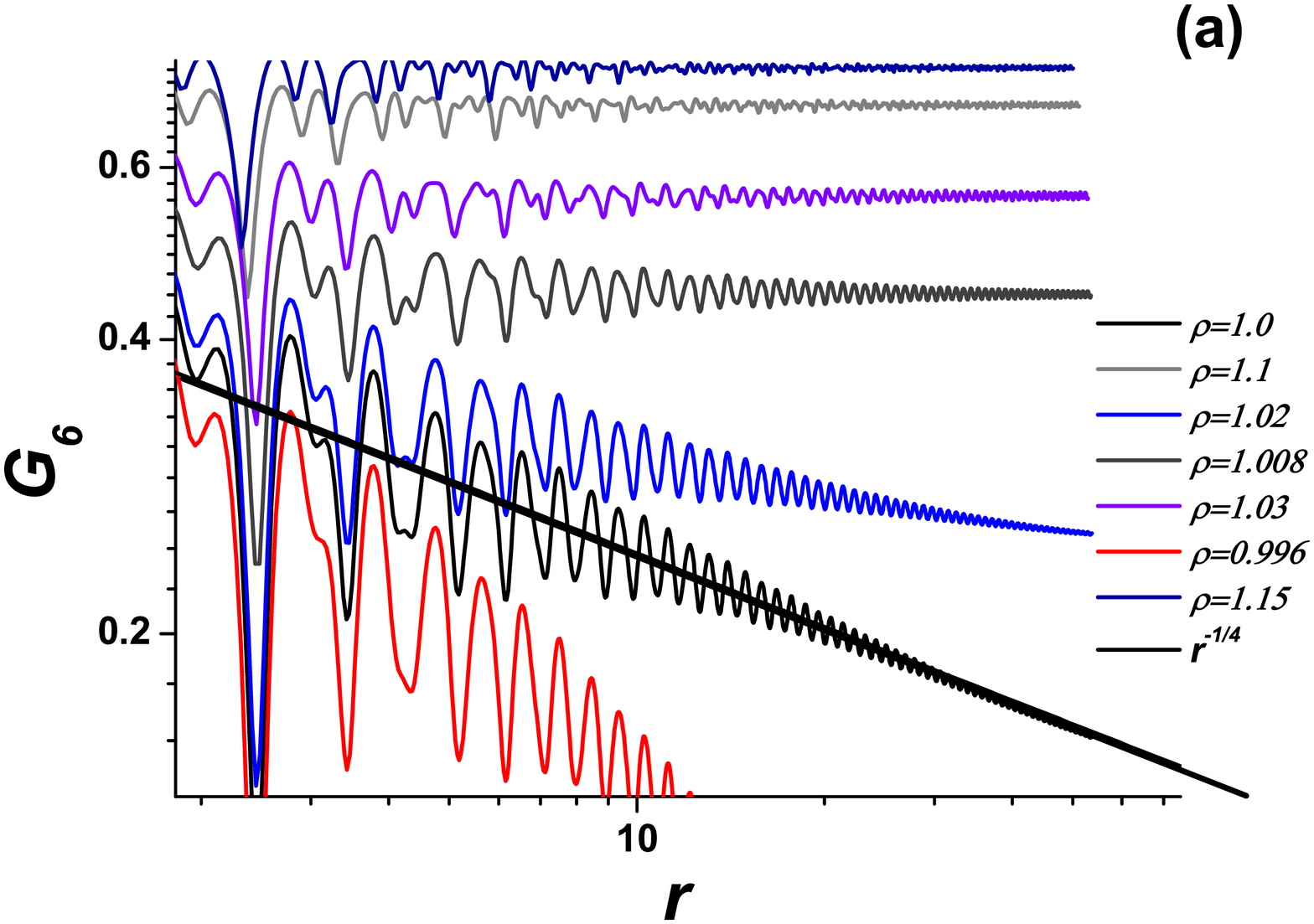}
\includegraphics[width=8cm, height=8cm]{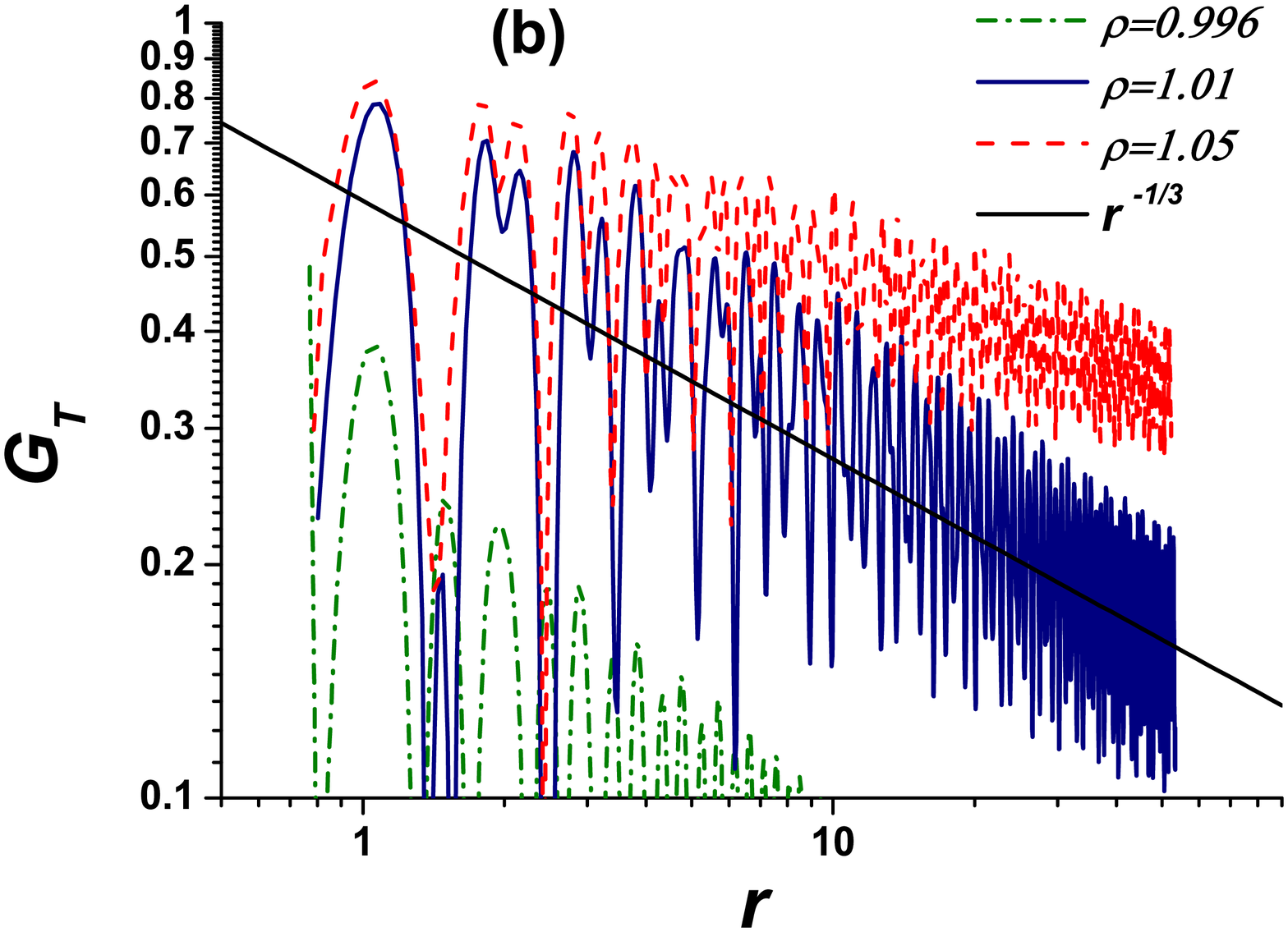}

\caption{\label{g6gt} (a) The orientational correlation functions
of the system with $n=12$. (b) The translational correlation
functions of the system with $n=12$.}
\end{figure}

Fig. \ref{g6gt} shows the correlation functions of the
orientational and translational order parameters of the system
without random pinning. They also confirm the presence of the
hexatic phase in the system. One can see that at high densities
the orientional correlation function does not decay, which
corresponds to the presence of long-range orientational order in
the system. At the same time the translational correlation
function decays algebraically. At low densities both orientational
and translational correlation functions demonstrate exponential
decay, i.e. isotropic liquid is observed.
The critical value of translational correlation function exponent
$\eta_T=1/3$ is reached at $\rho_{sh}=1.014$. This density is
higher than the first-order transition density obtained using the
Maxwell construction for the Mayer-Wood loop. The Maxwell
construction gives $\rho_l=0.998$ for the liquid density of the
liquid-hexatic transition and $\rho_{lh}=1.006$ for the density of
the hexatic phase. The limit of stability of the hexatic phase
$\eta_6=1/4$ is reached at $\rho_h=1.0$ which is inside the
Mayer-Wood loop. These densities are in good agreement with the
ones reported in \cite{foh4} ($\rho_l=0.998$, $\rho_{lh}=1.005$
and $\rho_{sh}=1.015$). Our results demonstrate that melting
occurs via the BK scenario.

The same type of calculations is performed for the system with
random pinning. As it was already mentioned, random pinning does
not influence orientational ordering, but it strongly affects the
translational order in the system. As a result, the limit of
stability of the hexatic phase shifts to greater densities, where
the crystal is stable in the pure system
\cite{ufn1,nel_dis1,nel_dis2,dis3,dis4}.

\begin{figure}
\includegraphics[width=8cm, height=8cm]{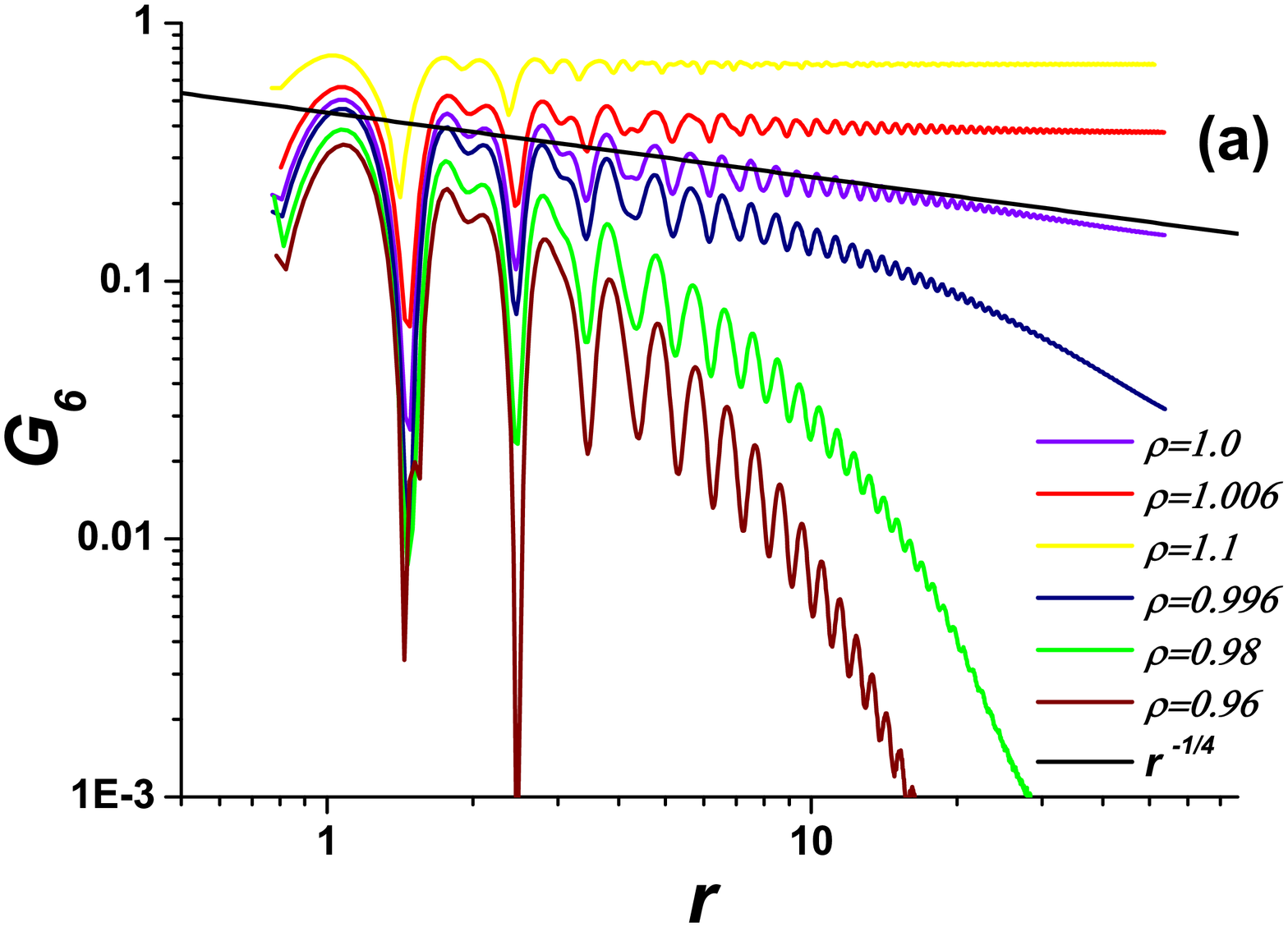}
\includegraphics[width=8cm, height=8cm]{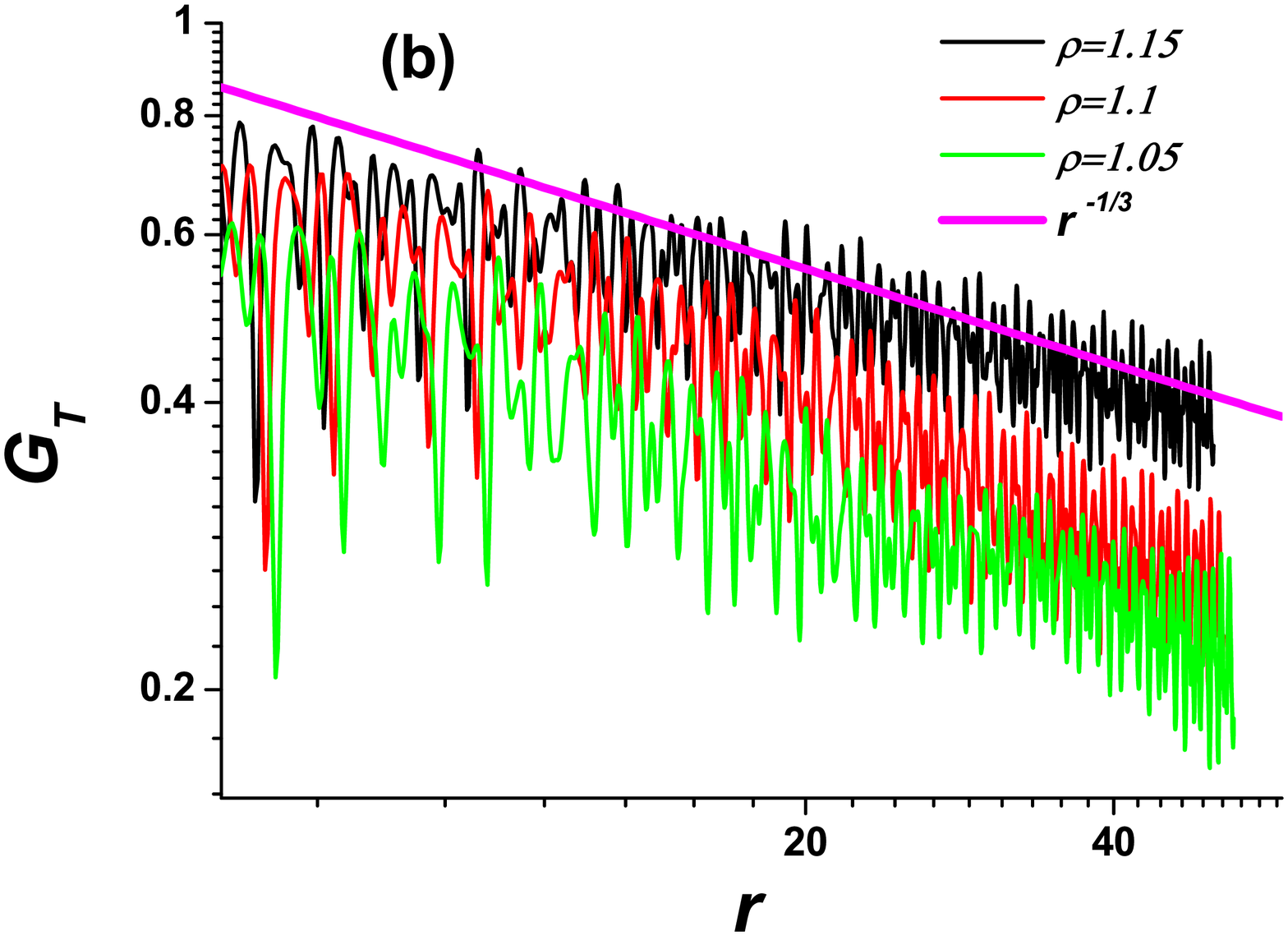}

\caption{\label{g6gt-pin} (a) The orientational correlation
functions of the system with $n=12$ with random pinning. (b) The
translational correlation functions of the system with $n=12$ with
random pinning.}
\end{figure}

Orientional correlation function $G_6$ and translational
correlation function $G_T$ of the system with random pinning are
shown in Fig. \ref{g6gt-pin}. As it was expected random pinning
does not change the location of the boundary between the hexatic
phase and isotropic liquid. However, the boundary between the
crystal and hexatic phases moved to higher densities of up to
$\rho_{sh}=1.15$. One can also see that the translational
correlation function of the system with random pinning
demonstrates a kink. At small distances random pinning does not
substantially affect local ordering. At the same time random
pinning does influence the asymptotic behavior of the
translational correlation functions. Because of this the region
after the kink should be used to determine the limit of stability
of the crystal phase.

\begin{figure}
\includegraphics[width=8cm, height=8cm]{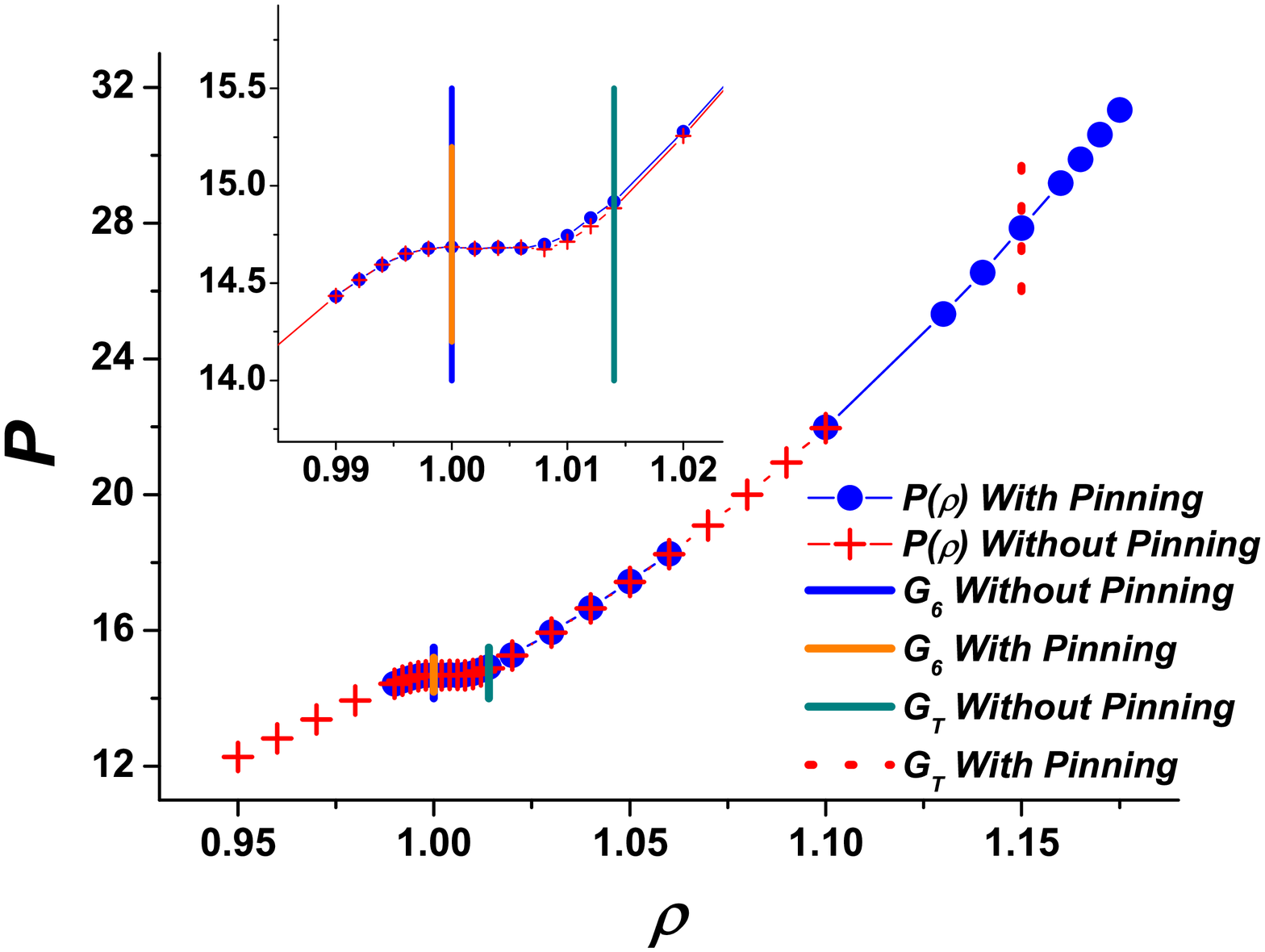}

\caption{\label{pb-12} The equation of state of the system with
$n=12$ with phase boundaries obtained from the $G_6$ and $G_T$
criteria. The inset enlarges the Mayer-Wood loop.}
\end{figure}

The phase boundaries obtained from $G_6$ and $G_T$ are placed on
the equation of state (Fig. \ref{pb-12}). One can see that random
pinning does not influence the transition from the hexatic phase
to isotropic liquid. At the same time the boundary between the
crystal and hexatic phases moved to much higher densities. This
observation allows us to improve our understanding of the system
melting scenario. Since the Mayer-Wood loop is observed, one can
be sure that a first-order phase transition takes place in the
system. However, in the pure system the stability region width of
the hexatic phase is very small and one can easily "miss"
\hspace{0.1cm} it. In the presence of random pinning the width of
the hexatic phase region becomes substantial and one can clearly
see that first-order phase transition corresponds to the
transition from the liquid to hexatic phases. No peculiarities are
observed at the crystal/hexatic phase boundary. This is especially
obvious in the system with random pinning since here this boundary
is far from first-order phase transition and does not in the least
affect the equation of state behavior. Therefore, this transition
is a continuous one. Our calculations confirm that the system
melts via the BK scenario.

In order to be more sure of the location where the phase boundary
of solid to hexatic transition is, we employ one more method based
on the elastic properties of a 2D system. According to the BKTHNY
theory the crystal transforms into the hexatic phase at
temperature $T_m$ which is determined by the BKT equation
\cite{ufn1,b1,kosthoul73,halpnel2}:
\begin{equation}
  k_BT_m=\frac{K}{16\pi}. \label{tm}
\end{equation}
Here $K$ is the reduced Young
modulus and $k_B$ is the Boltzmann
constant.

The reduced Young modulus of a
triangular crystal can be calculates as
\begin{equation}
  K=\frac{a_0^2}{k_B T}\frac{4 \mu (\mu + \lambda)}{2 \mu +
  \lambda},
\end{equation}
where  $\mu$ and $\lambda$ are usual Lame coefficients. $\mu$ has
the meaning of a shear modulus.

In order to calculate the Young modulus we calculate shear modulus
$\mu$ by the method proposed in Ref. \cite{broughton}. In this
method the system is strained. As a result a non-diagonal
component of pressure proportional to the shear modulus appears:

\begin{equation}
  P_{xy}=\mu u_{xy}+O(u_{xy}^2),
\end{equation}
where $u_{xy}$ is strain. Lame coefficient $\lambda$ can be
calculated if the bulk modulus is also known:

\begin{equation}
  B=\left( \frac{\partial P}{\partial \rho} \right)_T=\lambda +
  \mu.
\end{equation}

\begin{figure}
\includegraphics[width=8cm, height=8cm]{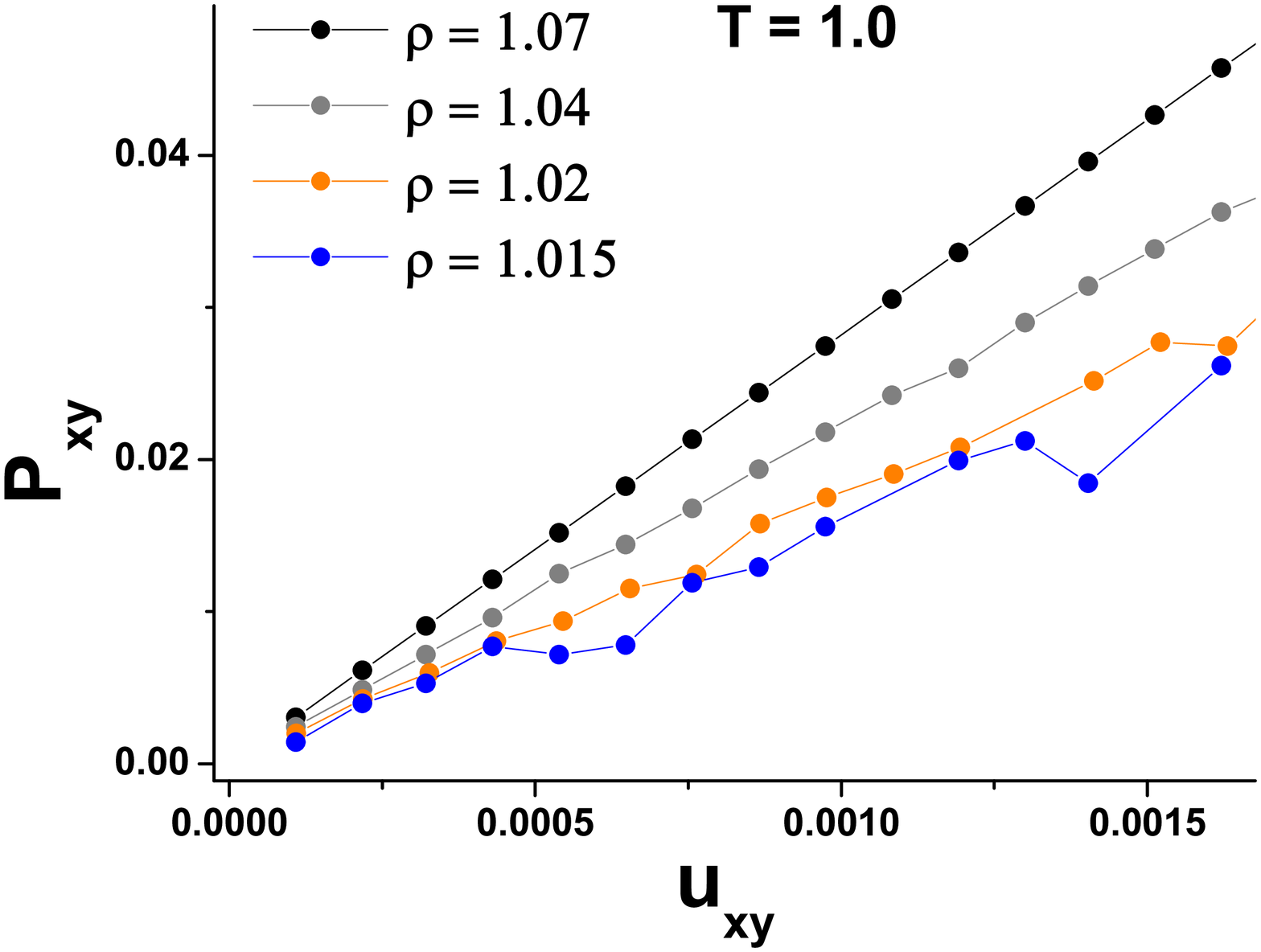}

\caption{\label{mu12} The non-diagonal component of pressure as a
function of applied strain in the pure system with $n=12$.}
\end{figure}

Fig. \ref{mu12} shows examples of $P_{xy}$ as a function of
strain. One can see that at high densities far from the transition
point the dependence is characterized by strong linearity. The
accuracy of calculations becomes worse as transition to the
hexatic phase is approached.

\begin{figure}
\includegraphics[width=8cm, height=8cm]{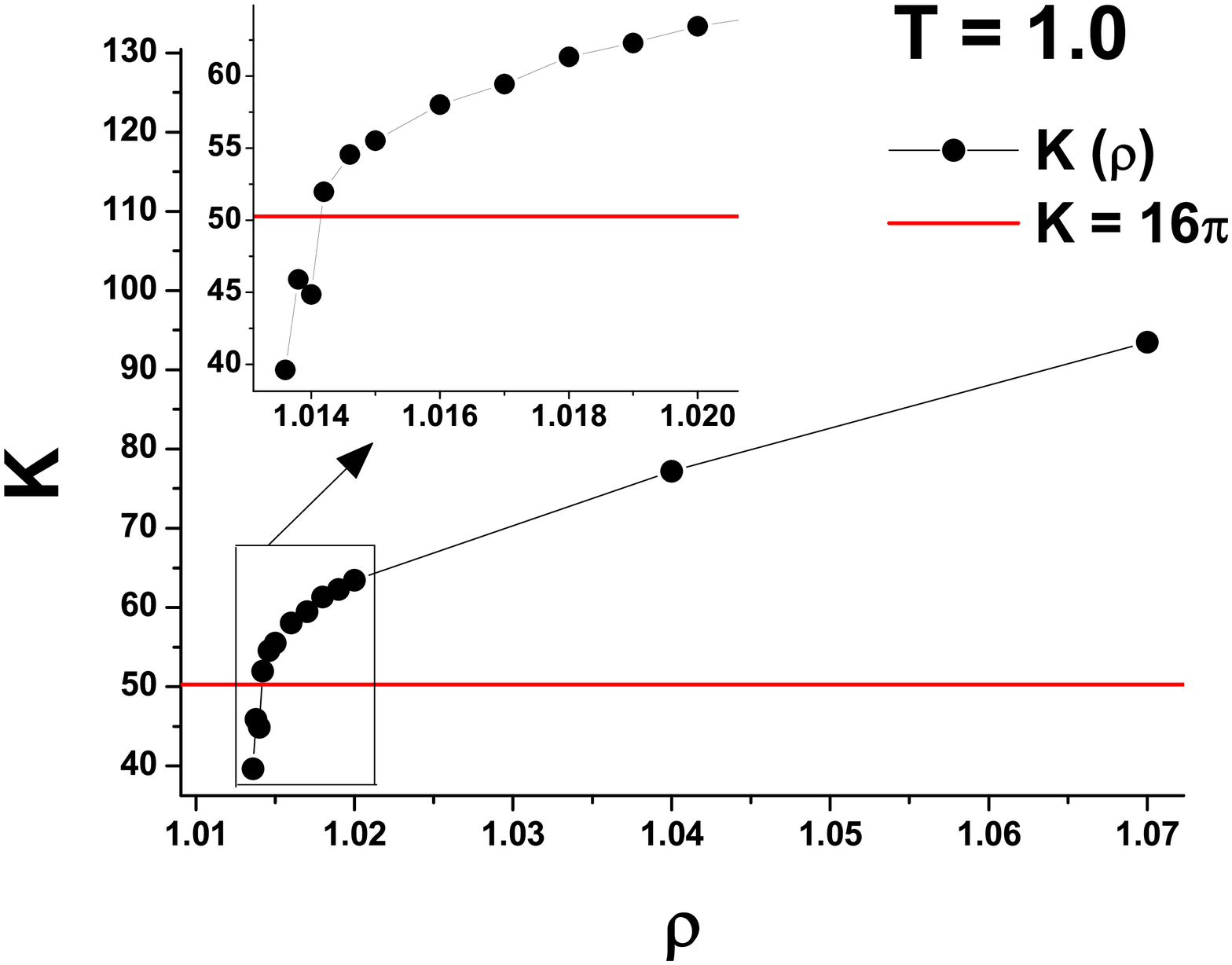}

\caption{\label{young} The Young modulus of the pure system with
$n=12$.}
\end{figure}

The resulting Young modulus is shown in Fig. \ref{young}. One can
see that the Young modulus reaches magnitude $16 \pi$ at
$\rho=1.014$ which is in good agreement with the results obtained
from $G_T$. However, such calculations of the Young modulus can
suffer from finite size effects. In order to remove these effects
one needs to use renormalization group equations. These equations
have the following forms:

\begin{equation}
\frac{d K^{-1}(l)}{dl}=\frac{3}{2} \pi y^2(l) e^{\frac{K(l)}{8
\pi}}I_0\left(\frac{K(l)}{8 \pi}\right)-\frac{3}{4} \pi y^2(l) e^{\frac{K(l)}{8
\pi}}I_1\left(\frac{K(l)}{8 \pi}\right)
\end{equation}

and

\begin{equation}
\frac{d y(l)}{d l}=\left( 2 - \frac{K(l)}{8\pi}\right)y(l)+2 \pi y^2(l)
e^{\frac{K(l)}{16\pi}}I_0\left(\frac{K(l)}{8\pi}\right),
\end{equation}
where $l$ is a scaling variable, $y$ is fugacity and $I_0$ and
$I_1$ are Bessel functions. Fugacity can be calculated as follows

\begin{equation}
 y=e^{-E_c/k_BT},
\end{equation}
where $E_c$ is the energy of the dislocation core. It can be
calculated from the probability of defect formation
\cite{binder1}:

\begin{equation}
   p_d=\frac{16 \pi \surd 3 \pi ^2}{K- 8 \pi}I_0\left(\frac{K}{8
   \pi}\right)e^{\frac{K}{8 \pi}}e^{-\frac{2E_c}{k_BT}}.
\end{equation}

\begin{figure}
\includegraphics[width=8cm, height=8cm]{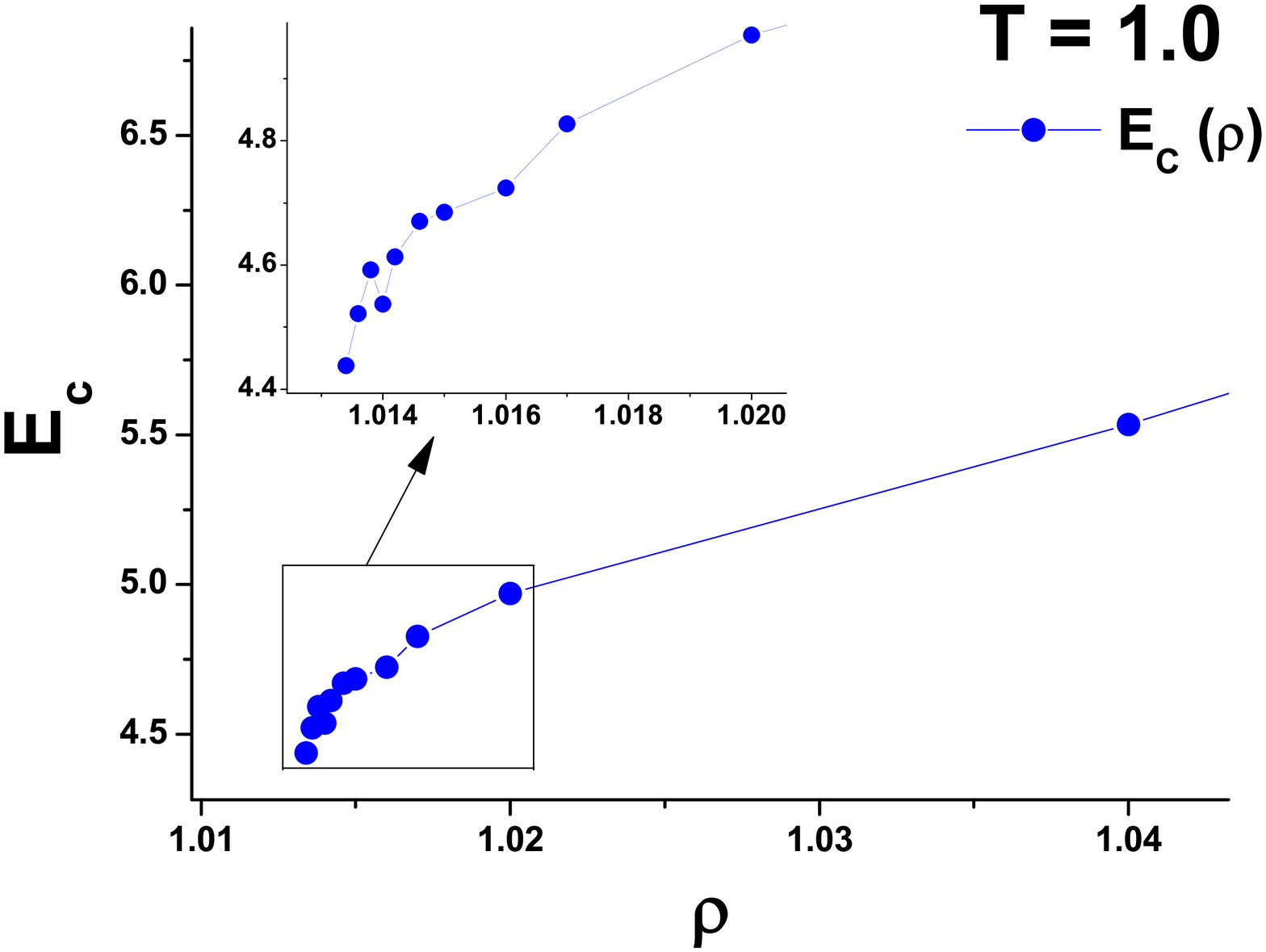}

\caption{\label{coreen} The dislocation core energy of the pure system with $n=12$.}
\end{figure}

Dislocation core energy is shown in Fig. \ref{coreen}. In Refs.
\cite{chui83,ryzhovJETP} it was shown that in the case of low
dislocation core energies the system should melt via first-order
phase transition. As it was mentioned in the Introduction, the
characteristic scale of dislocation core energy was given in Ref.
\cite{chui83} (see also Refs. \cite{ufn1,ryzhovJETP}), where it
was shown that transition went from strong first-order to weak
first-order for dislocation core energy $E_c$ less than $2.84
T_m$, where $T_m$ is taken from Eq. (\ref{tm}). Along isotherm
$T=1.0$ the critical core energy is $E_c=2.84$. One can see in
Fig. \ref{coreen} that the core energy obtained in our work is
considerably higher, therefore a continuous transition between the
crystal and hexatic phases is expected.

\begin{figure}
\includegraphics[width=8cm, height=8cm]{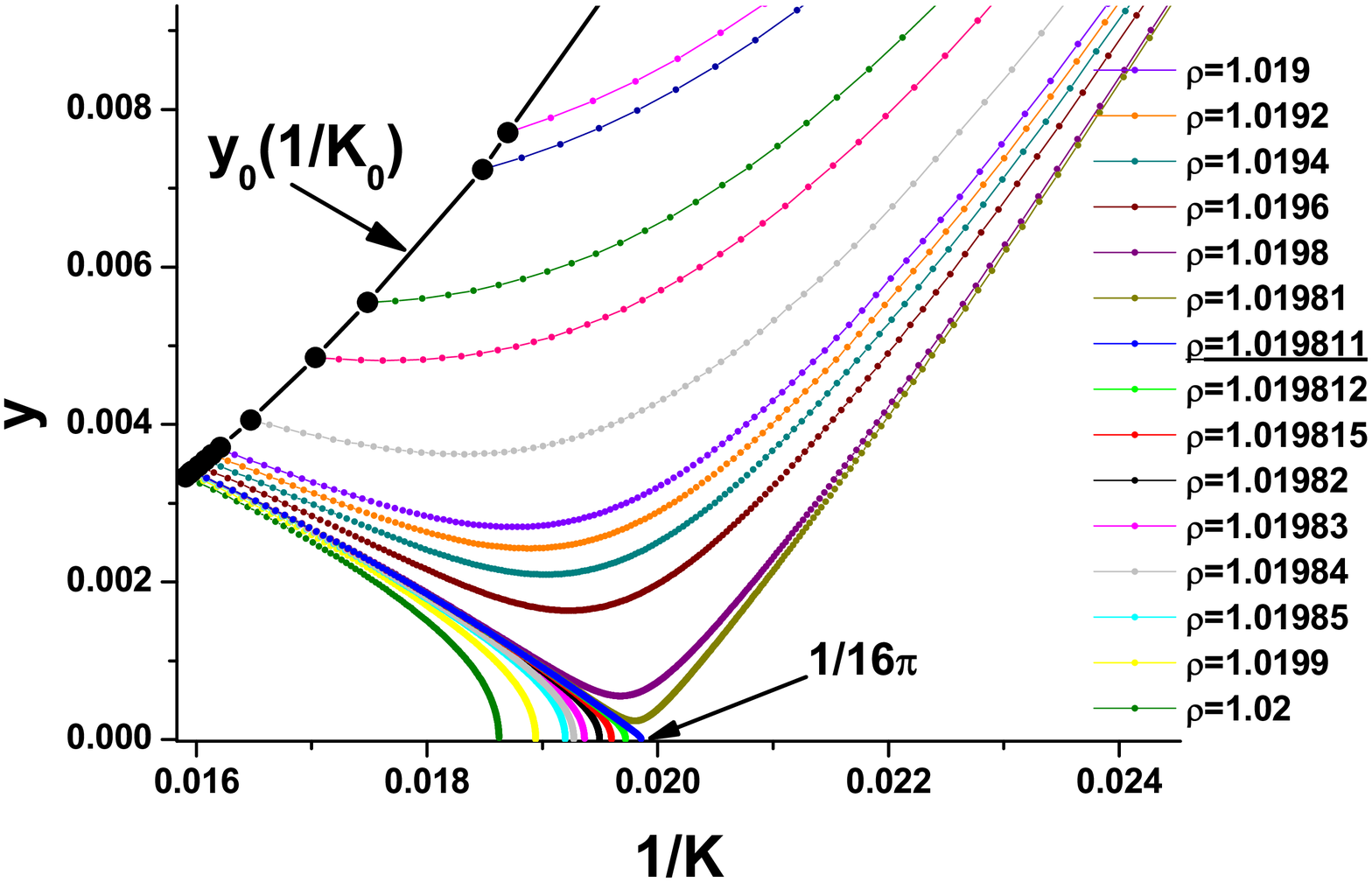}

\caption{\label{rg} The solution of the renormalization group
equations for the pure system with $n=12$.}
\end{figure}

Fig. \ref{rg} shows a solution of the renormalization group
equations for the pure system with $n=12$. The densities where $y
\rightarrow 0$ as $l \rightarrow \infty$ correspond to the
crystalline phase while the ones where $y \rightarrow \infty$ are
in the hexatic phase. From Fig. \ref{rg} one can see that
transition from crystal to hexatic takes place at
$\rho_{sh}=1.02$. This is somewhat higher than the results from
the $G_T$ criterion, however, the difference is within $0.5 \%$,
therefore the results can be considered as being in perfect
agreement.

In order to ensure that the hexatic phase is an ordered liquid we
perform calculations of the diffusion coefficient. Importantly,
the hexatic phase was observed in some publications (see, for
instance, \cite{prest1,foh1,foh2,we1,dfrt4}), but the region of
stability of this phase was always vanishingly small which
hindered investigation of hexatic phase properties. Because of
this it appears reasonable to study the properties of the hexatic
phase in a system with random pinning where it demonstrates a
large enough region of stability.

\begin{figure}
\includegraphics[width=8cm, height=8cm]{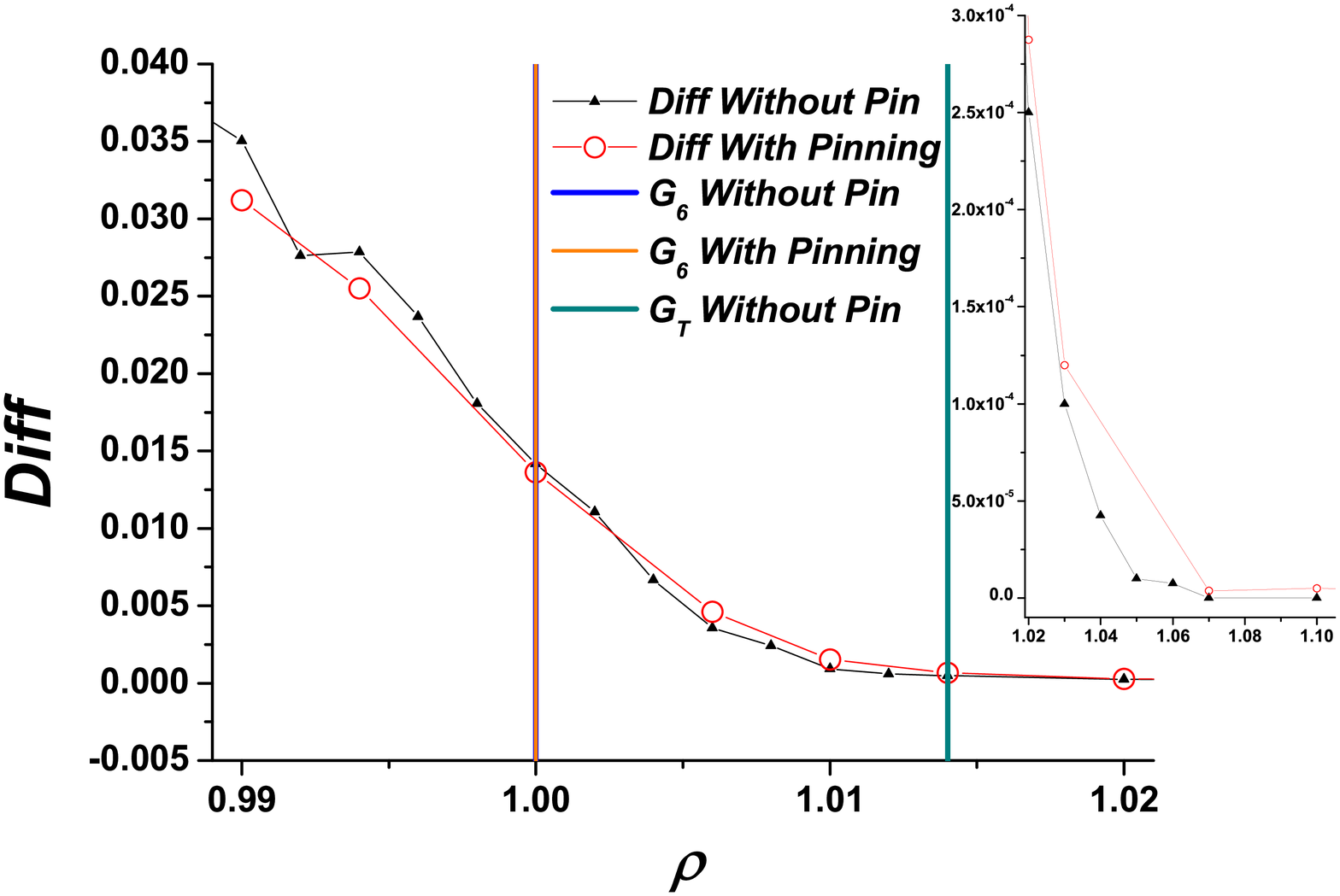}

\caption{\label{fig:fig10} The diffusion coefficient in the pure
system and in the system with random pinning with $n=12$.}
\end{figure}

Comparing the diffusion coefficients of a pure system with those
of a system with random pinning one can see that in the latter
case the finite value of the diffusion coefficient is preserved up
to density $\rho_{sh}=1.15$ (see the inset of Fig.
\ref{fig:fig10}) which corresponds to the boundary of
crystal-hexatic transition from the $G_T$ criterion. In isotropic
liquid the diffusion coefficient is about $11 \%$ higher in the
pure system compared to the system with random pinning.

The same type of calculations is performed for the pure system with
$n=1024$. Importantly, because of a very large n the system is
very anharmonic. Therefore, sufficient statistics are required to
obtain precise data. We calculated equations of state for
densities from $\rho_{min}=0.878$ up to $\rho_{max}=0.94$ and
$T=1.0$. We found the Mayer-Wood loop in this region. In the
region of the loop we averaged the system over 20 different
replicas (10 replicas of 20000 particles, 5 replicas of 45000
particles and 5 replicas of 8000 particles). The replicas differed
in the initial velocities of system particles. Only averaging over
20 replicas allowed us to remove noise from the results.

\begin{figure}
\includegraphics[width=8cm, height=8cm]{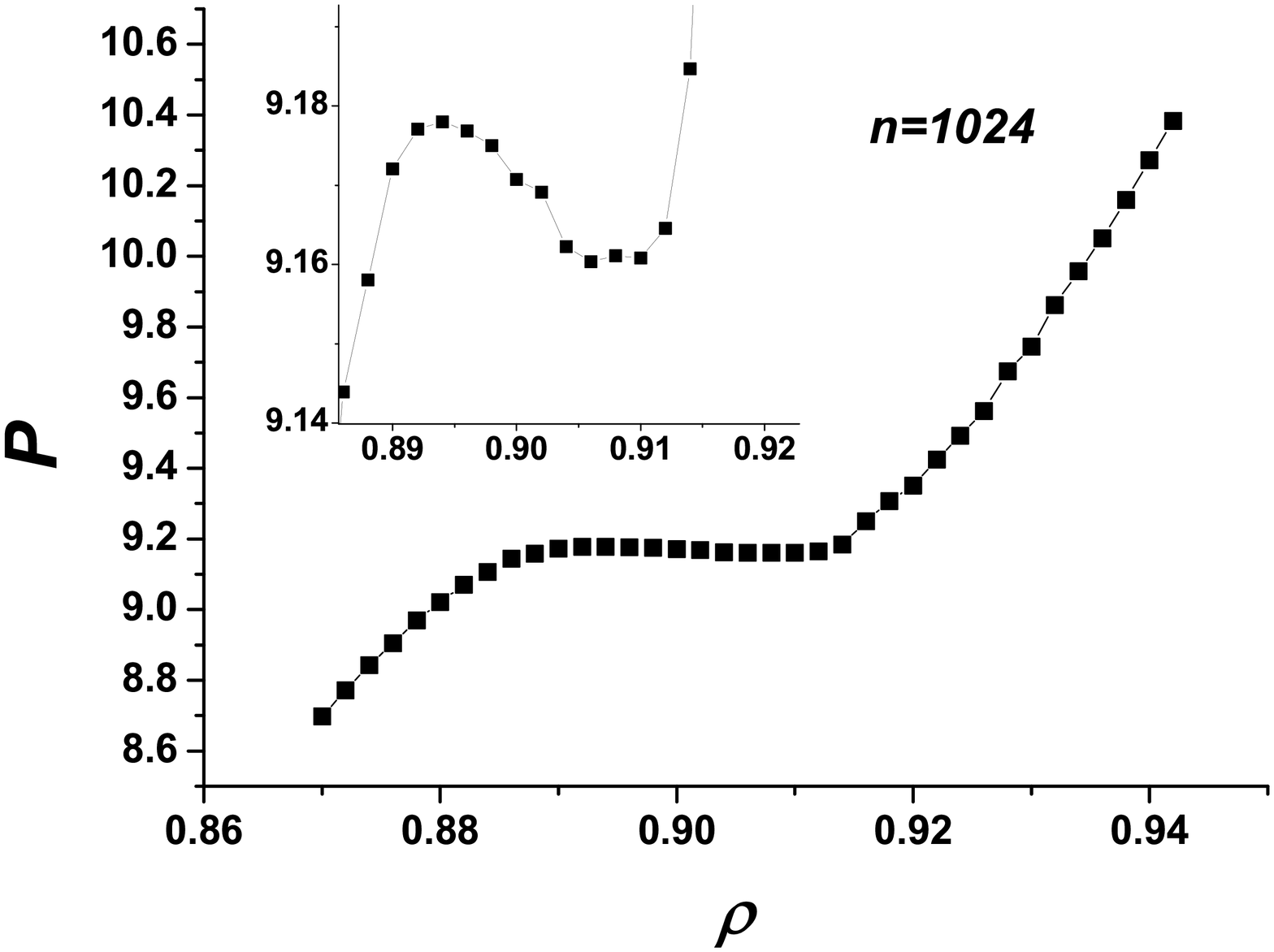}

\caption{\label{eos-1024} The equation of state of the pure system with
$n=1024$. The inset enlarges the region of the Mayer-Wood loop.}
\end{figure}

\begin{figure}
\includegraphics[width=8cm, height=8cm]{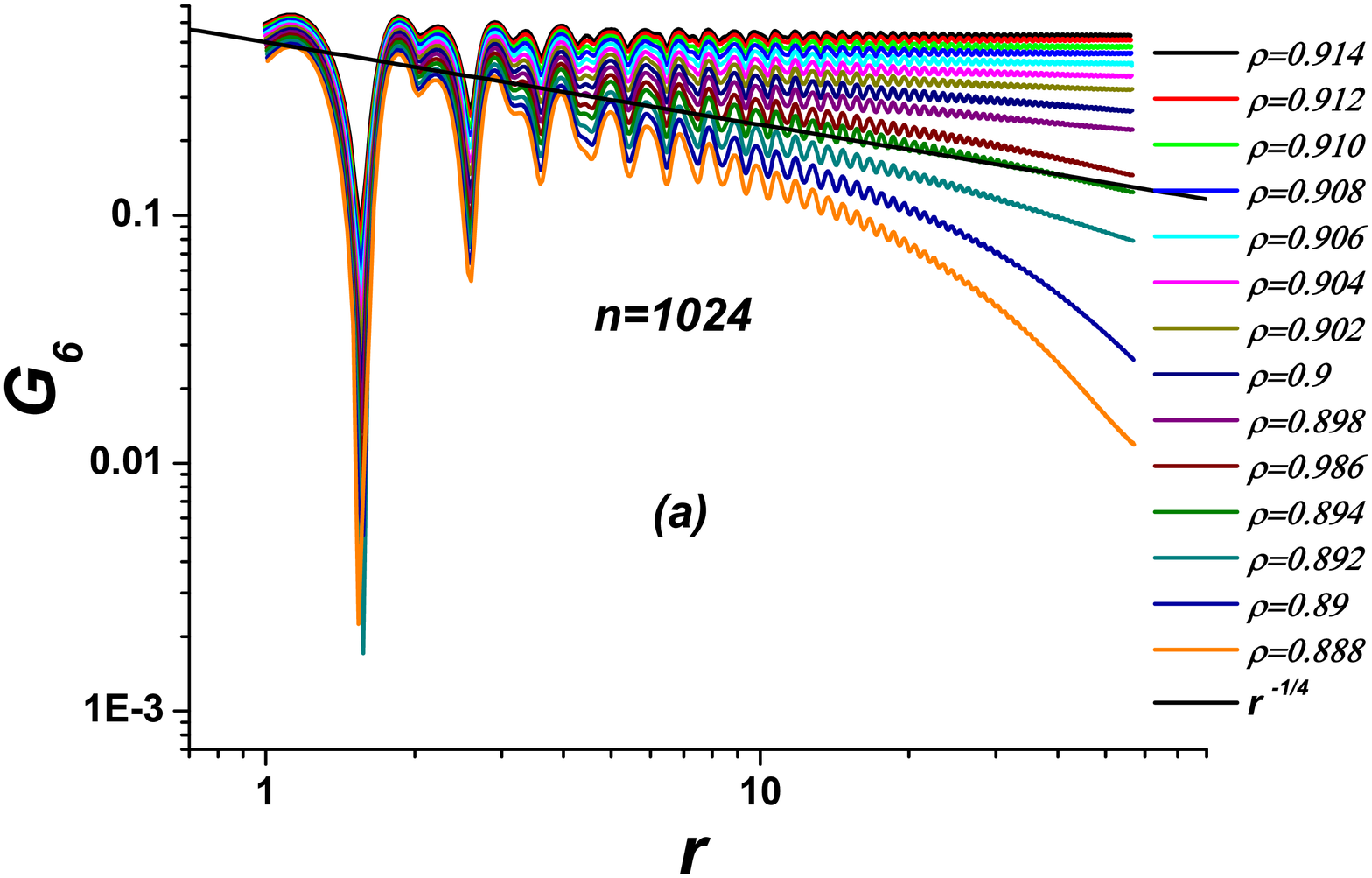}
\includegraphics[width=8cm, height=8cm]{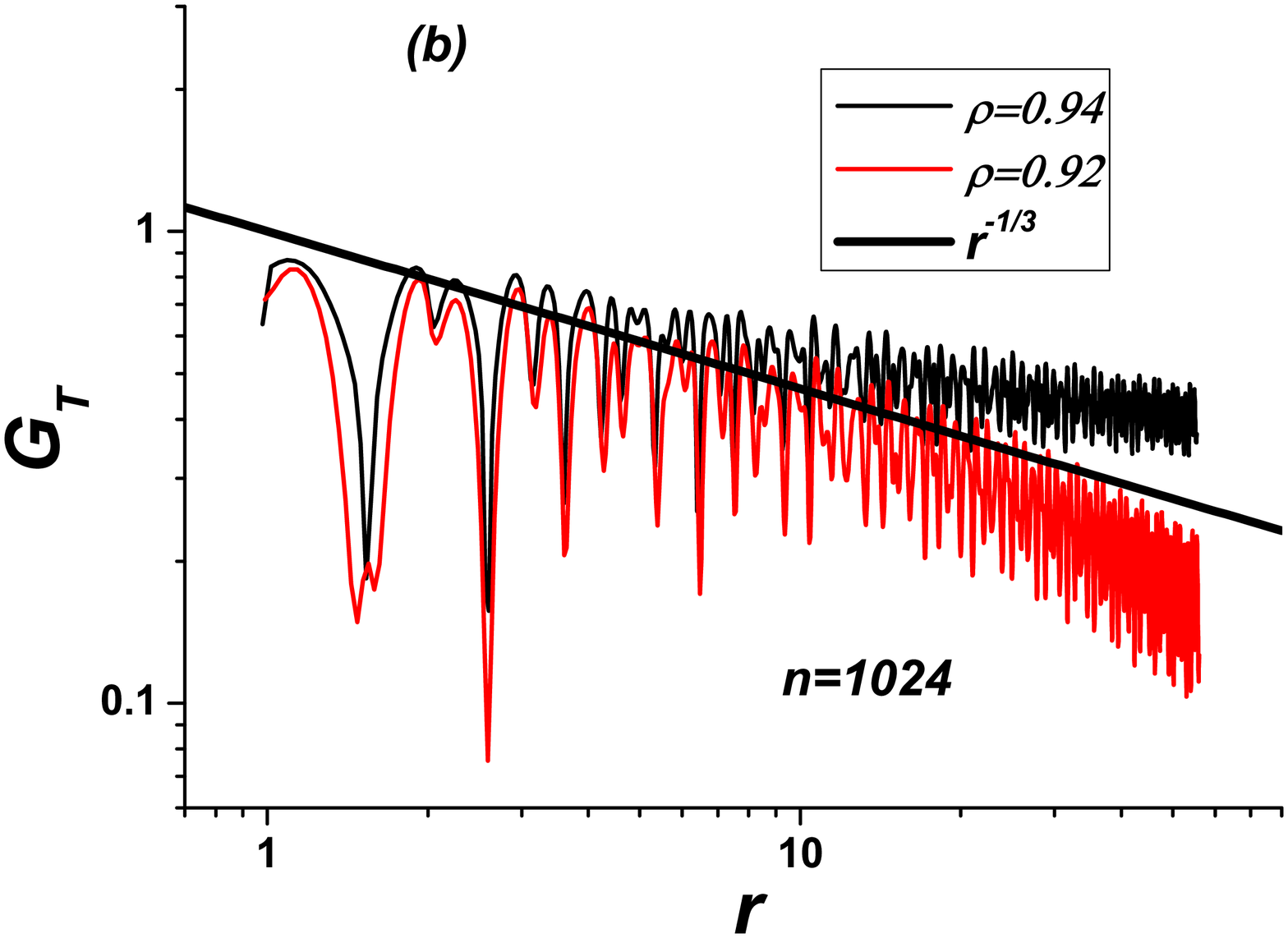}

\caption{\label{g6gt-1024} (a) The orientational correlation
functions of the pure system with $n=1024$. (b) The translational
correlation functions of the pure system with $n=1024$.}
\end{figure}

Fig. \ref{eos-1024} shows the equation of state of the system. One
can see the Mayer-Wood loop. From the Maxwell construction we
obtained a region of loop existence from density $\rho_l=0.89$ and
$\rho_{lh}=0.91$. To identify phase stability regions the
orientational and translational correlation functions were used
(Fig. \ref{g6gt-1024}). One can see that the stability limit of
the hexatic phase with respect to liquid is at $\rho_h=0.894$
(Fig. \ref{g6gt-1024}(a)) while that of crystal with respect to
the hexatic phase is $\rho_s=0.93$ (Fig. \ref{g6gt-1024}(b)).
These results also correspond to the BK scenario of melting.
Therefore, the "hardness" \hspace{0.1cm} of the potential does not
affect the melting scenario. However, the width of the hexatic
phase in the system with $n=1024$ appears to be larger.

In conclusion, we have performed a computer simulation study of a
soft-disk system with $n=12$ and $n=1024$. Pure systems and
systems with random pinning were studied. It was found that the
presence of random pinning made the stability region of the
hexatic phase larger. However, random pinning did not change the
melting scenario. Moreover, random pinning did not influence the
orientational ordering in the system, while it strongly affected
the translational order which is clearly seen from the behavior of
the orientational and translational order parameters and their
correlation functions. For the first time the diffusion
coefficient of the hexatic phase was calculated for this system.
We showed that the results for solid to hexatic transition
obtained from the $G_T$ criterion, Young modulus calculation and
the solution of renormalization group equations were in perfect
agreement. It was shown that the diffusion coefficient became
finite at the transition point of the crystal to hexatic phases.
In isotropic liquid the diffusion coefficient of the system with
pinning was slightly less than in the case of the pure system.

Importantly, in the vicinity of crystal to hexatic continuous
transition we did not observe any peculiarities of the equation of
state which was not very informative for the correct determination
of the melting scenario. That is why we used other criteria such
as the behavior of the orientational and translational order
parameters, their correlation functions, diffusion coefficient as
well as solving renormalization group equations.

Finally, it is noteworthy that the nature of first-order
transition from the hexatic phase to isotropic liquid is still not
understood. It should be noted that the
Berezinskii-Kosterlitz-Thouless transition can be made first-order
by reducing core energy $E_c$ of a corresponding topological
defect (disclination) below some critical value
\cite{f1,f2,f3,f4,f5,f6,f7}.

It should be also noted that our results confirm the melting scenario proposed in Ref. \cite{foh4} and contradict the recent article \cite{kor}. This contradiction will be discussed in further publications.

\begin{acknowledgments}
The authors are grateful to E.E. Tareyeva, and V.V.
Brazhkin for valuable discussions. This work was carried out using
computing resources of the federal collective usage center
"Complex for simulation and data processing for mega-science
facilities" at NRC "Kurchatov Institute", http://ckp.nrcki.ru, and
supercomputers at Joint Supercomputer Center of the Russian
Academy of Sciences (JSCC RAS). The work was supported by the
Russian Foundation for Basic Research (Grant No 17-52-53014 (VNR)
and 18-02-00981 (YDF)).

\end{acknowledgments}


\end{document}